\documentclass{aa}  

\usepackage{graphicx}
\usepackage{txfonts}
\usepackage{lipsum}
\usepackage{subcaption}         
\usepackage{lscape}             
\usepackage{placeins}           
                                
\usepackage[colorlinks,allcolors=blue]{hyperref}
\usepackage{breakurl}
\usepackage{stackengine}
\usepackage{amsmath}

\newcommand {\beq}{\begin {eqnarray}}
\newcommand {\eeq}{\end {eqnarray}}

\begin{document}

\title{Ultra-luminous X-ray pulsars as sources of TeV neutrinos}
\titlerunning{Ultra-luminous X-ray pulsars as sources of TeV neutrinos}

\author{L.~Ducci
\inst{1,2}
\and
E.~Perinati
\inst{1}
\and
P.~Romano
\inst{2}
\and
S.~Vercellone
\inst{2}
\and
M.~Niko{\l}ajuk
\inst{3}
\and
A.~Santangelo
\inst{1}
\and
M.~Sasaki
\inst{4}
}

\institute{Institut f\"ur Astronomie und Astrophysik, Kepler Center for Astro and Particle Physics, University of Tuebingen, Sand 1, 72076 T\"ubingen, Germany\\
\email{ducci@astro.uni-tuebingen.de}
\and
INAF -- Osservatorio Astronomico di Brera, via Bianchi 46, 23807 Merate (LC), Italy
\and
Faculty of Physics, University of Bia{\l}ystok, ul. K. Cio{\l}kowskiego 1L, 15-245 Bia{\l}ystok, Poland
\and
Dr. Karl Remeis Observatory, Erlangen Centre for Astroparticle Physics, Friedrich-Alexander-Universit\"at Erlangen-N\"urnberg,
Sternwartstr. 7, 96049 Bamberg, Germany
}

\abstract{We explored the expected properties of the neutrino emission from accreting neutron stars in X-ray binaries using numerical simulations. The simulations are based on a model in which neutrinos are produced by the decay of charged pions and kaons, generated in inelastic collisions between protons accelerated up to TeV energies in the magnetosphere of a magnetized ($B\sim 10^{12}$~G) neutron star and protons of the accretion disc. Our results show that this process can produce strong neutrino emission up to a few tens of TeV when the X-ray luminosity is above $\sim 10^{39}$~erg~s$^{-1}$, as in ultra-luminous X-ray (ULX) pulsars. We show that neutrinos from a transient Galactic ULX pulsar with $L_{\rm x} \approx 5\times 10^{39}$~erg~s$^{-1}$ can be detected with kilometre-scale detectors such as IceCube if the source is within about 3--4~kpc. We also derived an upper limit on the neutrino flux from the Galactic ULX pulsar Swift J0243.6+6124 using IceCube data, a result that has not been previously reported. Our findings establish a new benchmark for future astrophysical neutrino observations, critical for interpreting data from current and upcoming instruments with significantly improved sensitivity.}

  \keywords{accretion -- stars: neutron -- X-rays: binaries -- neutrinos -- X-rays:individual (Swift J0243.6+6124)}

  \maketitle  
%

\section{Introduction}

One of the major breakthroughs in the context of multi-messenger astrophysics of the last decade
was the discovery of TeV-PeV neutrinos of astrophysical origin
by the IceCube neutrino observatory \citep{Aartsen13, 2013Sci...342E...1I, Aartsen14a}.
Although the dominant contribution is from extragalactic sources,
neutrino emission has recently been identified from the Galactic plane
at a $4.5\sigma$ level of significance, which contributes $\sim 6$ to 13\% of the all-sky astrophysical flux at 30~TeV \citep{2023Sci...380.1338I}.
This excess is likely dominated by diffuse emission from our Galaxy, but a fraction of it
could be linked to a population of unresolved point sources \citep[e.g.][]{2023Sci...380.1338I, Vecchiotti23}.
Some of these sources could be X-ray binaries (XRBs), which have long been considered
candidates for neutrino emission \citep[e.g.][]{Berezinskii85, Gaisser85, Auriemma86}.
XRBs are composed of a compact object (a neutron star or a black hole) and, typically, a non-degenerate companion star.
X-rays are the result of the accretion of the mass lost by the companion star onto the compact object.
The gravitationally captured mass sometimes forms an accretion disc.
Some XRBs are persistent X-ray sources but most display variability, sometimes bright X-ray outbursts
that can last several weeks followed by long periods of low X-ray luminosity or quiescence \citep[for two recent reviews on XRBs, see, for example,][]{Kretschmar19, Sazonov20}.

Over the years, three primary mechanisms for neutrino production in XRBs have been identified.
The first and second mechanisms involve interactions of accelerated hadrons. In the first case, high-energy protons accelerated within the XRB, such as in jets, relativistic winds, or neutron star magnetospheres, collide with protons in environments such as stellar winds, accretion discs, or the jets themselves. These collisions produce primarily pions, which decay into gamma rays and neutrinos. This mechanism has been proposed for microquasars, gamma-ray binaries, and other XRBs \citep[e.g.][]{Anchordoqui03, Romero03, Bednarek05, Christiansen06, Torres07, Reynoso08, Neronov09, Sahakyan14, Kosmas23, Peretti24}.
The second mechanism occurs when accelerated protons interact with photon fields from the accretion disc, companion star, or neutron star surface. Here, pion production is dominated by the $\Delta$-resonance process, with subsequent decays generating neutrinos and gamma rays. The proton--photon interaction is particularly relevant for microquasars and accreting pulsars \citep{Levinson01, Distefano02, Bednarek09}. Neutrinos produced via these mechanisms typically lie in the GeV--PeV energy range.
A third mechanism arises in accretion columns of bright, transient XRBs. High temperatures in these regions are expected to lead to electron--positron pair production, with neutrino emission occurring via pair annihilation ($e^- + e^+ \rightarrow \nu + \bar{\nu}$). Additionally, neutrino synchrotron emission ($e + B \rightarrow e + \nu + \bar{\nu}$) contributes to the total neutrino luminosity, though its contribution is comparatively minor. These processes produce neutrinos in the MeV energy band \citep{Mushtukov18, Asthana23, Mushtukov25}.

In hadronic scenarios within XRBs, gamma-ray and neutrino emission are intrinsically linked because both originate from the decay of neutral and charged pions produced in high-energy hadron interactions. 
However, unlike neutrinos, gamma rays are susceptible to absorption within the dense photon environments of XRBs, primarily through pair production. This process occurs when high-energy gamma rays interact with lower-energy photons from the companion star (UV radiation) and the accretion disc (X-rays). The significant absorption of gamma rays can lead to a quenched or substantially altered observed gamma-ray emission \citep[see e.g.][]{Torres07, Neronov09, Aharonian11, Ducci23}.
In contrast, neutrinos interact very weakly with matter and radiation and can therefore escape the dense astrophysical environments of XRBs without significant absorption. Consequently, neutrinos can provide a more direct and undistorted view of the fundamental hadronic acceleration processes occurring in these systems.

Neutrino emission from XRBs has been actively investigated using detectors such as IceCube \citep{Abbasi22}
and the Astronomy with a Neutrino Telescope and Abyss environmental RESearch project (ANTARES; \citealt{Albert17}).
While intriguing excesses in neutrino event rates have been reported near sources such as Cyg X-3 in IceCube data \citep{Koljonen23} and in Baikal Deep Underwater Neutrino Telescope or Baikal-Gigaton Volume Detector observations \citep{Allakhverdyan23}, no statistically significant neutrino detection has been confirmed to date. Current theoretical models predict neutrino fluxes from XRBs that remain below the sensitivity thresholds of existing detectors \citep{Abbasi22, Albert17}.

In this work, we extend the hadronic neutrino production mechanism proposed by \citet{Anchordoqui03} to the broader range of mass accretion rates (or equivalently, X-ray luminosities) characteristic of highly magnetized ($B\sim 10^{12}$~G) accretion disc-fed pulsars.
This mechanism relies on the hypothesis of an electrostatic gap forming in the neutron star magnetosphere. While theoretically motivated, the existence, stability, and detailed behaviour of such gaps under accretion conditions remain unconfirmed observationally and lack validation via self-consistent numerical magnetohydrodynamic or particle-in-cell simulations \citep[see, for example,][ and references therein]{Cerutti15}.
Our refined predictions for system-specific neutrino fluxes will enable better comparisons with future observations from next-generation detectors such as IceCube-Gen2 \citep{Grant19} and KM3NeT \citep{Aiello24}, whose enhanced sensitivity will provide critical tests of astrophysical neutrino models.
    
The paper is organized as follows: Section \ref{sect.model} presents the theoretical framework and model assumptions. Section \ref{sect.simu} details the numerical simulations adopted to derive our results, which are described in Sect. \ref{sect.results}. Concluding remarks are given in Sect. \ref{sect.conclusions}.

\section{Model synopsis} \label{sect.model}

Our calculations and simulations build on the model proposed by \citet{Anchordoqui03},
which itself extends the framework developed by \citet{Cheng89}.
The central idea involves electrostatic gaps forming in the neutron star magnetosphere due to differential rotation between the star
and the inner accretion disc. These gaps accelerate protons to energies that can exceed 100~TeV.
When these protons collide with the accretion disc, they trigger hadronic showers.
\citet{Cheng89} considered the gamma-ray emission from the decay of neutral pions produced by the proton--proton collisions.
\citet{Anchordoqui03} investigated the  high-energy neutrino production via charged pion decay and, to a lesser extent, charged kaons
that decay into neutrinos. 

The model by \citet{Anchordoqui03} assumes a strongly shielded gap, where the magnetospheric region is devoid of
X-ray photons produced by the accretion, maximizing the potential drop:
\begin{equation} \label{eq:DV_strong}
     \Delta V_{\rm strong} \approx 4\times 10^{14} \beta^{-5/2} \left( \frac{M}{M_\odot}\right)^{1/7} R_6^{-4/7} L_{\rm x,37}^{5/7} B_{12}^{-3/7} \mbox{ V\ ,}
\end{equation}
where $L_{\rm x,37}$ is the X-ray luminosity produced by the accretion in units of $10^{37}$\,erg\,s$^{-1}$,
$M$  is the mass of the neutron star, $R_6$  is its radius in units of $10^6$\,cm,
$B_{12}$ is the magnetic field at the neutron star surface (in units of $10^{12}$\,G), 
$\beta=2R_0/R_{\rm A}$, where
\begin{eqnarray}
  R_0      & = &  1.35\gamma_0^{2/7}\eta^{4/7}R_{\rm A} \mbox{\ ,}  \label{eq:R0}\\
  R_{\rm A} & = & \mu^{4/7} \dot{M}^{-2/7} (2 G M)^{-1/7} \mbox{\ .} \label{eq:RA}
\end{eqnarray}
Here $R_0$ is defined according to the work by \citet{Wang96}, $R_{\rm A}$ is the Alfv\'en radius,
$\gamma_0=-B_{\phi_0}/B_{z_0}\approx 1$ is the magnetic pitch angle,
$\eta$ is the screening factor (a partially screened disc is assumed, $\eta=0.1$)\footnote{$\eta$ takes into account the screening effects of the currents induced in the surface of the accretion disc.},
$\mu=BR_{\rm NS}^3/2$ is the dipolar magnetic moment ($R_{\rm NS}$ is the radius of the neutron star), and $\dot{M}$ is the mass accretion rate.
In the model of \citet{Anchordoqui03}, the electrostatic gap is situated over the accretion disc from $R_0$ up to a distance of about $R_{\rm A}$.

\citet{Cheng91} and \citet{Cheng92} later introduced the weakly shielded gap scenario to improve the physical realism of the model, a feature not considered by \citet{Anchordoqui03}.
In the weakly shielded gap scenario, X-ray photons from the accretion column penetrate the gap, producing $e^\pm$ pairs that reduce the potential:
\begin{equation} \label{eq:DV_weak}
  \Delta V_{\rm weak} \approx 6\times 10^{11} L_{\rm x,37} \left( \frac{E_{\rm x}}{\mbox{1\,keV}}\right)^{-1} R_6^{-1} B_{12}^{-1/2} \mbox{ V\ ,}
\end{equation}
where $E_{\rm x}$ is the characteristic energy of the X-ray photons \citep{Cheng91, Cheng92}.
Weak shielding lowers proton acceleration energies compared to the strongly shielded case.

The proton current through the gap, critical for determining the neutrino flux, is given by \citep{Cheng89}
\begin{equation} \label{eq:Jmax}
 J_{\rm p}/e \approx 3.1\times 10^{33} \beta^{-2} \left(\frac{M}{M_\odot} \right)^{-2/7} R_6^{1/7} L_{x,37}^{4/7} B_{12}^{-1/7} \mbox{\,s}^{-1} \mbox{\ .}
\end{equation}
To simulate the hadronic processes initiated by proton--proton collisions, \citet{Anchordoqui03} employed DPMJET-II, a Monte Carlo event generator designed
for modelling high-energy hadronic interactions using Gribov-Regge theory \citep{Ranft1995},
and Geant4, a comprehensive toolkit for simulating the passage of particles through matter,
widely used in high-energy physics and applicable to astrophysical phenomena \citep{Agostinelli03}\footnote{For more information: \url{https://geant4.web.cern.ch/}}.
While DPMJET generated the initial proton-nucleon interactions and secondary particle spectra,
Geant4 tracked the subsequent electromagnetic and hadronic shower development within the disc, including pion decays.
Their simulations revealed a power-law neutrino spectrum, predicting that systems like the Be/XRB A0535+26 could emit detectable neutrino signals,
highlighting the potential for multi-wavelength astrophysics to probe hadronic processes in extreme environments.

\section{Numerical simulations} \label{sect.simu}

In our study, we focused on the weakly shielded gap scenario, which was not addressed by \citet{Anchordoqui03}, to complement and enhance their work.
We assumed a mono-energetic proton beam striking the accretion disc perpendicularly near $R_{\rm A}$, with the proton energy determined by Eq. \ref{eq:DV_weak} and their flux by Eq. \ref{eq:Jmax}. For each simulation run, we generated between $10^4$ and $10^6$ protons in the beam and subsequently scaled the results using Eq. \ref{eq:Jmax}.
The accretion disc was modelled as a cylindrical volume with height $2h$, filled with a gas of ionized protons of density $n_{\rm p}$. The density of the disc and height at $\sim R_{\rm A}$ were derived using the disc structure equations from \citet{ss73}. For the radiation pressure dominated cases (the so-called zone A), in our simulations with $L_{\rm x} \gtrsim 10^{38}$~erg~s$^{-1}$  we adopted the scale-height formula reported in \citet{Bykov22} and \citet{Mushtukov17}: $(2h/R)_{\rm m} = 0.1 (L_{\rm x}/10^{39}\mbox{\,erg\,s}^{-1}) (R_{\rm NS}/10^6\mbox{\,cm}) (M_{\rm NS}/M_\odot)^{-1} (R_{\rm m}/10^8\mbox{\,cm})^{-1}$. As pointed out in \citet[][ and references therein]{Mushtukov17}, for high values of the X-ray luminosity (higher than those assumed in our simulations), the accretion disc can become advection-dominated and the thickness given by the above formula can be overestimated. For $n_{\rm p}$, we referred to the numerical calculations by \citet{Inoue23} and \citet{Chashkina19}.

Within the disc, the interaction between the beam and target protons generates particle cascades, leading to the creation of various particle types. Our focus is on those that result in the formation of neutrinos. At the energies given to the beam protons, neutrinos are produced from the initial creation of $\pi^\pm$ and, to a lesser extent, $K^\pm$ mesons that emerge from the proton--proton collision and any subsequent hadronic interactions involving $\pi^\pm$ and $K^\pm$, with the target protons. Neutrinos arise from the decays of these mesons, the most probable of which are%
\begin{eqnarray}
  X^+ & \rightarrow & \mu^+ + \nu_\mu \nonumber \\
  X^- & \rightarrow & \mu^- + \bar{\nu}_\mu \nonumber,
\end{eqnarray}
where $X$ is $\pi$ or $K$.
Subsequently, muons can further decay, producing additional neutrinos:
\begin{eqnarray}
  \mu^+ & \rightarrow & e^+ + \nu_e + \bar{\nu}_\mu \nonumber \\
  \mu^- & \rightarrow & e^- + \bar{\nu}_e + \nu_\mu \nonumber. 
\end{eqnarray}
In conclusion, there will be a mix of muon and electron neutrinos. 

In addition to proton--proton collisions, proton--photon interactions represent another potential mechanism for meson and neutrino production in high-energy astrophysical environments.
In our study, we focused exclusively on proton--proton interactions for meson production for two reasons.
First, the energy threshold required for efficient meson production in proton--photon processes, particularly those involving interactions with accretion disc photons, is of the order of $10^2$~TeV \citep{Aharonian11, Neronov09}. Given that the majority of proton energies explored within our simulation framework reside below this threshold, the kinematic requirements for effective proton--proton interactions are often not met.
Second, the proton--photon cross-section is at least two orders of magnitude smaller than the proton--proton cross-section (e.g. \citealt{Reynoso08, Kelner08}).
Given the proton and photon densities in our simulations, the proton--proton channel dominates the proton--photon production process.

   \begin{figure*}
   \centering
   \includegraphics[width=\columnwidth+\columnwidth]{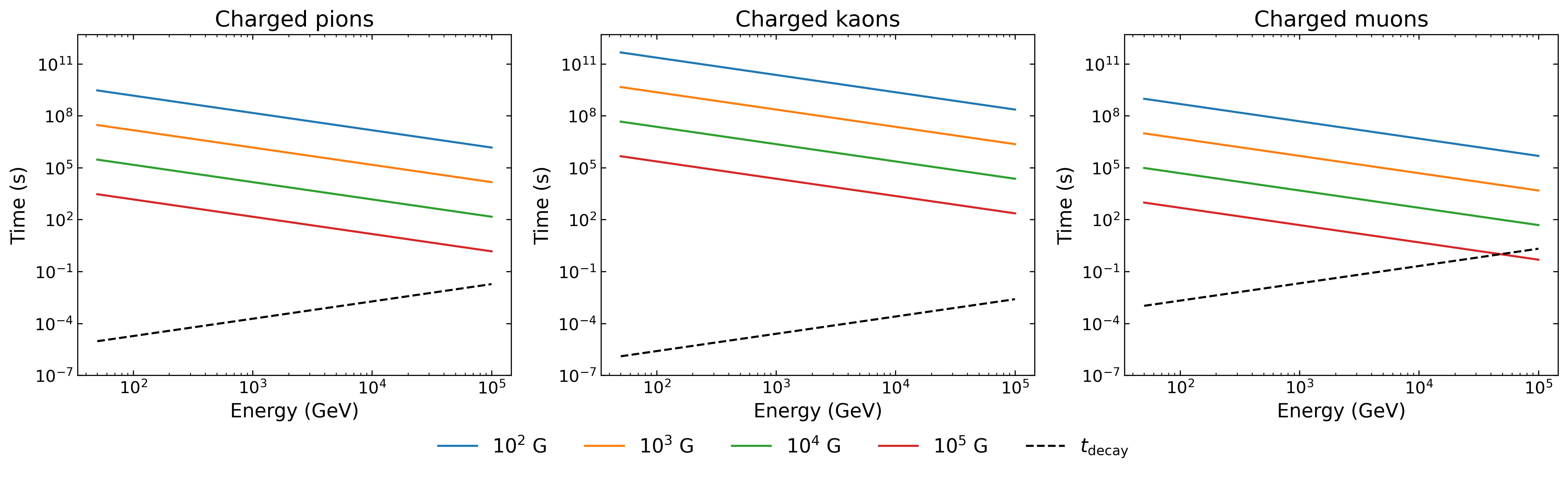}
   \caption{Synchrotron cooling times for pions, kaons, and muons, for different magnetic field strengths, as a function of energy, compared to their lifetimes in the accretion disc rest frame.}
   \label{tcool}
   \end{figure*}

We performed our simulations entirely using Geant4 (version 11.3.0). This choice reflects the steady advancements in the software over the past $\sim 22$ years, which have established the {\tt FTFP\_BERT} physics list\footnote{For more details, see \url{https://geant4.web.cern.ch/documentation/dev/plg_html/PhysicsListGuide/reference_PL/FTFP_BERT.html}} as the recommended option for high-energy physics applications, including hadronic interactions and decays. Nevertheless, we emphasize that ongoing validation studies remain critical to assess existing models and extend the physics implementations against other particle interaction/transport codes and experimental data (see e.g. \citealt{Allison16}).
{\tt FTFP\_BERT} is optimized for the GeV to TeV energy range and includes elastic, inelastic, and capture processes for the hadronic interactions.
For high-energy inelastic collisions ($\gtrsim 4$~GeV), the Fritiof (FTF) model governs the production of high-multiplicity secondary particles, notably charged pions and kaons.
These particles either undergo secondary interactions within the target (the accretion disc) or decay via electromagnetic/weak processes, producing muons and neutrinos. To quantify the neutrino flux and energy spectrum, Geant4 tracks the full decay chains of pions, kaons, and muons, while secondary hadronic interactions are modelled to account for cascade-driven modifications to the final state of the particle distributions.
To optimize computational efficiency, we stopped tracking particles once their energies fell below 50~GeV. This cutoff minimizes time-consuming low-energy cascade calculations while preserving fidelity to the GeV-TeV neutrino spectrum, the primary focus of astrophysical detectors like IceCube.

\begin{figure}
   \centering
   \includegraphics[width=\columnwidth]{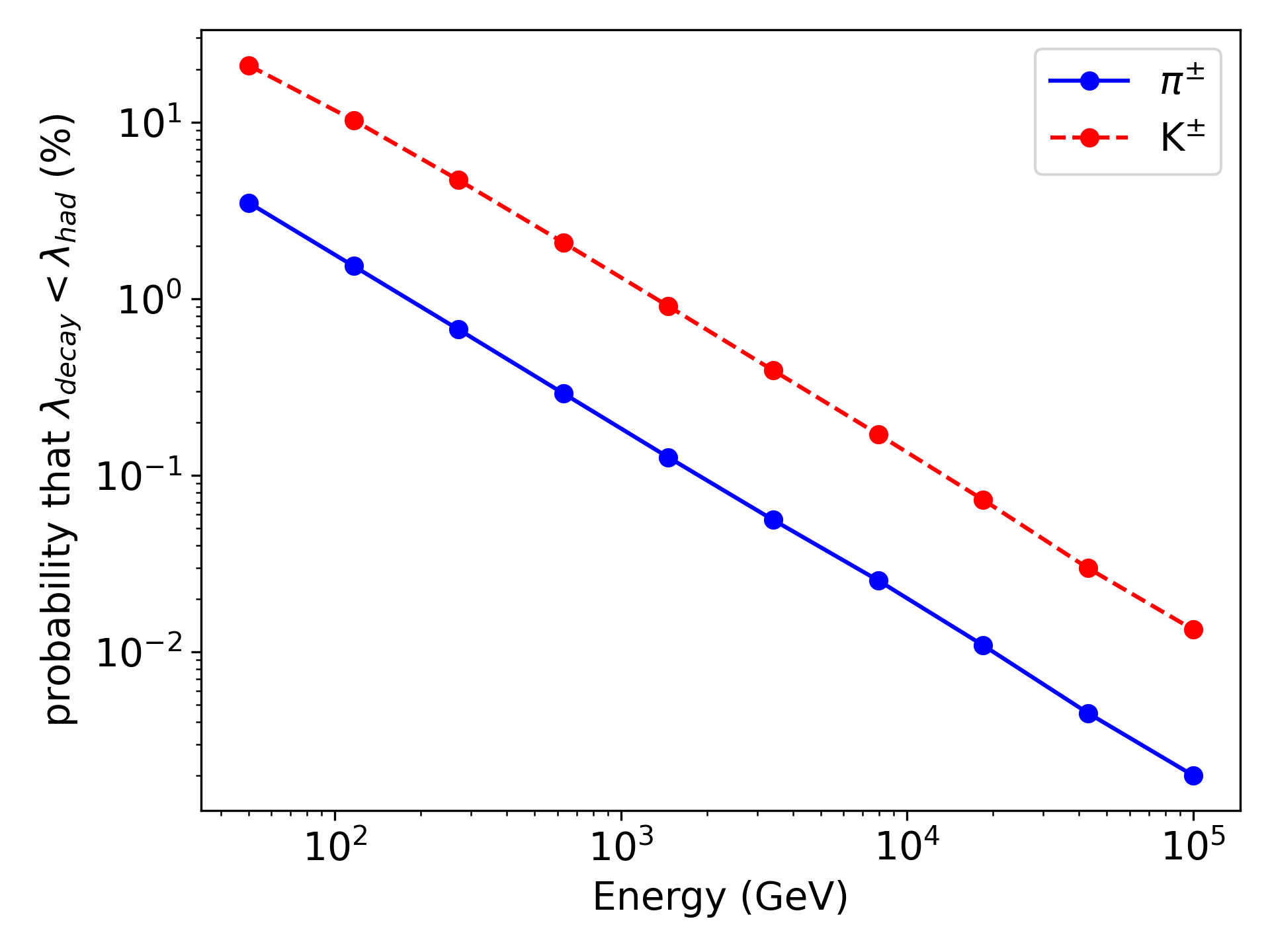}
   \caption{Probability that the mean free path for decay of $\pi^\pm$ and $K^\pm$ is less than the mean free path for inelastic collisions with protons of the accretion disc as a function of energy, assuming $n_{\rm p}=5\times 10^{21}$~cm$^{-3}$.}
   \label{lambda_decay_vs_inelastic}
   \end{figure}

The magnetic field strength at $R_{\rm A}$, given by $B(R_{\rm A}) = B_0 (R_{\rm NS}/R_{\rm A})^3$ (for a dipolar field with $B_0$ at the neutron star surface), is relatively low when typical values for accreting pulsars in high-mass XRBs (a sub-class of XRBs; see e.g. \citealt{vanParadijs98}) are used.
This makes synchrotron losses for $\pi^\pm$, $K^\pm$, and $\mu^\pm$ negligible in most cases, as also shown by \citet{Anchordoqui03}.
Figure \ref{tcool} shows that synchrotron losses only slightly affect muons at very high X-ray luminosities ($L_{\rm x}\gtrsim 10^{39}$~erg~s$^{-1}$), where $B(R_{\rm A}) \gtrsim 10^5$~G.
However, in these high mass accretion regimes, $n_{\rm p}$ is also very high. This strongly reduces the probability that muons precursors, $\pi^\pm$ and $K^\pm$,
decay before inelastic collisions with protons of the disc.
An example is shown in Fig. \ref{lambda_decay_vs_inelastic}, which shows the probability that $\pi^\pm$ and $K^\pm$
decay before undergoing inelastic collisions with protons in the disc (assuming $n_{\rm p}=5\times 10^{21}$~cm$^{-3}$).
Therefore, the magnetic field strength can be neglected in first approximation.

\section{Results} \label{sect.results}

\begin{table*}
\begin{center}
\caption{Summary of the simulations.}
\label{table simu}
\begin{tabular}{lcccccccc}
\hline
\hline
\noalign{\smallskip}
$L_{\rm x}$        &         $2h$       &        $n_{\rm p}$     &  $E_{\rm p}$  &      $J_{\rm p}/e$  &    $B(R_{\rm A})$ &$N_{\rm p\,sim}$  &  $\gamma$  &             $L_\nu$            \\
\hline
\noalign{\smallskip}
erg~s$^{-1}$      &          cm        &         cm$^{-3}$      &    TeV      &       s$^{-1}$      &        G        &               &                   &        GeV/s      \\
\noalign{\smallskip}
\hline
\noalign{\smallskip}
$5\times 10^{36}$ &  $6.3\times 10^6$  &  $2.4 \times 10^{19}$  &    0.11     & $3.0\times 10^{33}$ & $1.8\times 10^3$ &    $10^6$    & $-7.8 \pm  0.3$   & $(4.2 \pm 0.2) \times 10^{31}$ \\
\noalign{\smallskip}
$10^{37}$         &  $5.6\times 10^6$  &  $5.1 \times 10^{19}$  &    0.22     & $4.4\times 10^{33}$ & $3.3\times 10^3$ &    $10^5$    & $-5.3 \pm  0.1$   & $(1.6 \pm 0.04) \times 10^{33}$ \\
\noalign{\smallskip}
$5\times 10^{37}$ &  $3.0\times 10^6$  &  $4.0 \times 10^{20}$  &    1.12     & $1.1\times 10^{34}$ & $1.3\times 10^4$ &    $10^4$    & $-3.41 \pm  0.05$ & $(8.0 \pm 0.2) \times 10^{34}$ \\
\noalign{\smallskip}
$10^{38}$         &  $2.8\times 10^6$  &  $7.3 \times 10^{20}$  &    2.24     & $1.6\times 10^{34}$ & $2.3\times 10^4$ &    $10^4$    & $-3.13 \pm  0.04$ & $(1.96 \pm 0.04) \times 10^{35}$ \\
\noalign{\smallskip}
$10^{39}$         &  $4.5\times 10^6$  &  $5 \times 10^{21}$    &    22.4     & $6.1\times 10^{34}$ & $1.6\times 10^5$ &    $10^4$    & $-2.94 \pm  0.08$ & $(1.51 \pm 0.08) \times 10^{36}$ \\
\noalign{\smallskip}
$5\times 10^{39}$ &  $2.1\times 10^7$  &  $5 \times 10^{21}$    &   100.0     & $1.5\times 10^{35}$ & $6.7\times 10^5$ &    $10^4$    & $-2.88 \pm  0.04$ & $(1.15 \pm 0.04) \times 10^{37}$ \\
\noalign{\smallskip}
\hline
\end{tabular}
\end{center}
    {\small Notes. \emph{Input parameters:} X-ray luminosity ($L_{\rm x}$ ), accretion disc thickness ($2h$), proton density in the disc ($n_{\rm p}$),
      energy of the protons of the beam ($E_{\rm p}$), flux of protons of the beam ($J_{\rm p}/e$), magnetic field at $R_{\rm A}$ ($B(R_{\rm A})$) assuming $B_0=5\times 10^{12}$~G,
      number of protons of the beam simulated ($N_{\rm p\,sim}$).
      \emph{Output parameters:} Value of the slope of the power-law from the best fit of the simulated spectra ($\gamma$, where the power-law function is $N = KE^{\gamma}$), neutrino luminosity above 50~GeV from the simulated spectra ($L_\nu$).
      }
\end{table*}

   \begin{figure}
   \centering
   \includegraphics[width=\columnwidth]{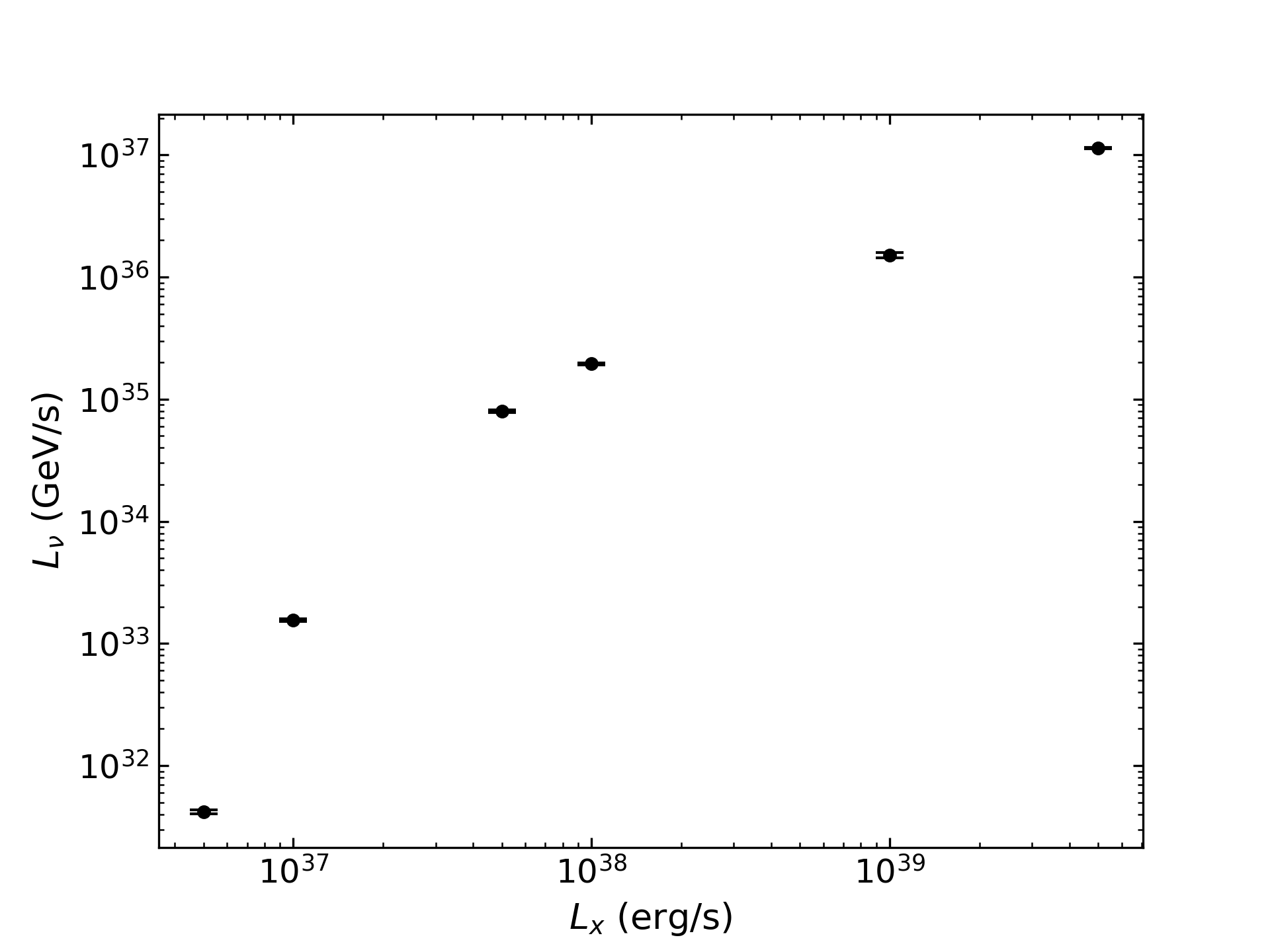}
   \caption{Neutrino luminosity as a function of X-ray luminosity for each of the simulated datasets (Table \ref{table simu}).}
   \label{Lnu_vs_Lx}
   \end{figure}

We explored the expected intensity and spectral properties of neutrinos from a hypothetical accreting pulsar in high-mass XRBs fed by an accretion disc across an X-ray luminosity range of $5\times 10^{36}$~erg~s$^{-1}$ to $5\times 10^{39}$~erg~s$^{-1}$. The neutron star was assumed to have a magnetic field of $B_0 = 5\times 10^{12}$~G, consistent with values typically observed in such systems \citep[e.g.][]{Staubert19}.
We assumed a neutron star with mass of $1.4$~M$_\odot$ and a radius of $12$~km.
We further assumed that the source remains in a persistent accretion state, with no inhibition of accretion at low luminosities \citep{Illarionov75}. The simulated neutrino spectra are parameterized using a power-law model to facilitate comparisons. For each simulation, the derived neutrino luminosities and power-law slopes are summarized in Table \ref{table simu}, alongside the corresponding input parameters. Figure \ref{Lnu_vs_Lx} presents the neutrino luminosity ($L_\nu$) as a function of the X-ray luminosity.
To account for neutrino flavour oscillations during propagation from the source to Earth, which redistribute the initial flux into approximately equal contributions of electron, muon, and tau neutrinos, the predicted neutrino luminosity was scaled by a factor of 1/3.
This adjustment accounts for the focus of detectors like IceCube on track-like events from muon neutrinos, which provide far better directional
precision for identifying astrophysical sources than cascade events produced by electron and tau neutrinos \citep{Aartsen14b, Aartsen20}.

The correlation between $L_\nu$ and $L_{\rm x}$ in the plot arises from the scaling of both the
proton flux (Eq. \ref{eq:Jmax}) and proton energy (Eq. \ref{eq:DV_weak}) with $L_{\rm x}$.
The sub-exponential scaling of $L_\nu$ with increasing $L_{\rm x}$ is caused by the higher proton density within the accretion disc. At these densities, charged mesons ($\pi^\pm$, $K^\pm$) with energies exceeding 50~GeV experience enhanced inelastic collisions prior to decay. These interactions induce energy redistribution into secondary particle cascades, suppressing the high-energy decay products (e.g. muons and neutrinos), leading to a flatter $L_\nu$ trend.

   \begin{figure}
   \centering
   \includegraphics[width=\columnwidth]{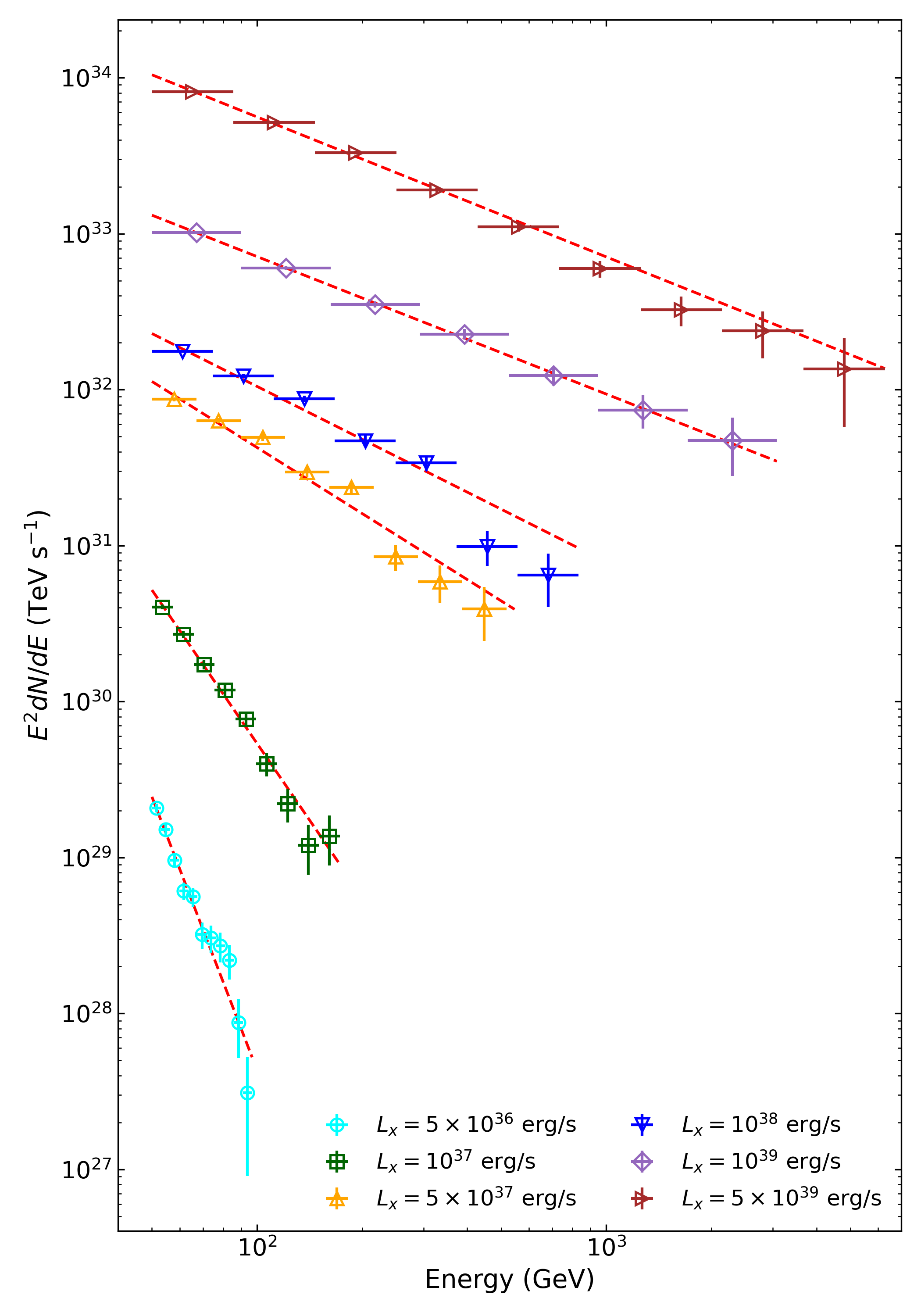}
   \caption{Simulated spectra for six different scenarios (Table \ref{table simu}), each fitted with a power-law model.}
   \label{spectra}
   \end{figure}

Figure \ref{spectra} shows the simulated spectra
where neutrino events are re-binned to a logarithmic scale
(refer to Table \ref{table simu}). As $L_{\rm x}$ increases, the spectra become harder.
This trend is primarily due to the increase in $E_{\rm p}$ with $L_{\rm x}$.
The variation in the power-law slope depends on a complex interplay among various interaction processes occurring within the accretion disc, competing with decays, as a function of the beam protons energy, target proton density, and disc thickness.

   \begin{figure}
   \centering
   \includegraphics[width=\columnwidth]{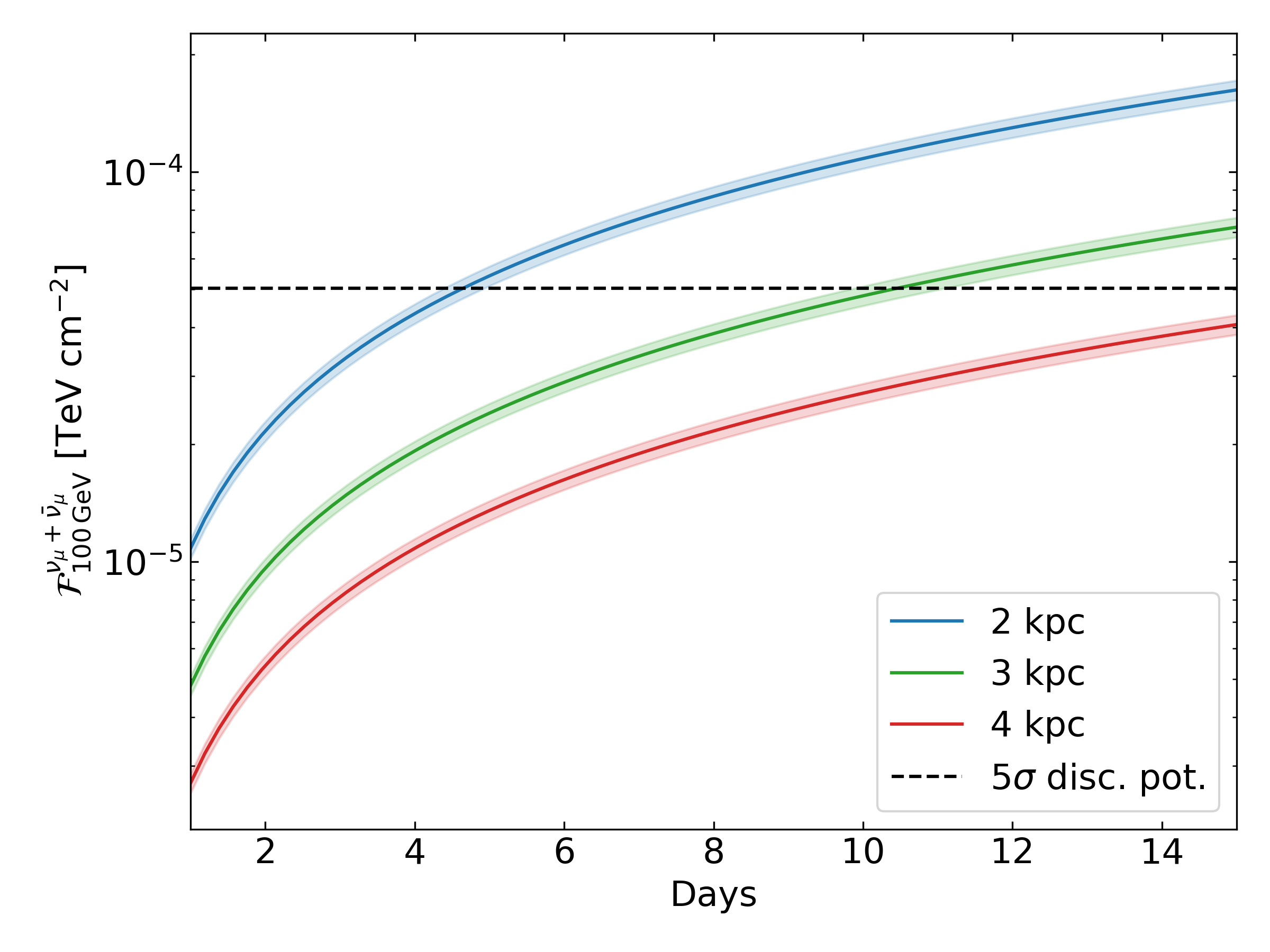}
   \caption{Fluence at 100~GeV of a hypothetical PULX with $L_{\rm X}=5\times 10^{39}$~erg~s$^{-1}$ as a function of the time interval during which the source maintains this luminosity level, for three different distances. The horizontal dashed line shows the $5\sigma$ discovery potential in IceCube.}
   \label{fluence}
   \end{figure}

Our simulations show that pulsars in binary systems with X-ray luminosities greater than $\sim 10^{39}$~erg~s$^{-1}$ are promising neutrino emitters within the model framework presented in this work.
Accreting pulsars in binary systems with such high luminosities are classified as ultra-luminous X-ray sources with pulsars (PULXs), and represent a sub-class of ultra-luminous X-ray sources (ULXs). PULXs are identified by the detection of X-ray pulsations, which confirm the presence of an accreting neutron star. In contrast, the broader ULX class includes systems thought to host stellar-mass black holes, intermediate-mass black holes, or neutron stars where pulsations have not yet been observed (e.g. \citealt{King23}). Approximately 1800 ULXs have been observed in nearby galaxies \citep{Walton22,Middleton17,Earnshaw19},
with only a few confirmed or strong candidate PULXs \citep{Bachetti14, Israel17a, Israel17b, Carpano18, Fuerst16, Sathyaprakash19, Rodriguez20, Pintore25}. In addition to these, three transient pulsars in the Magellanic Clouds, SMC~X-3 \citep{Tsygankov17}, RX~J0209.6-7427 \citep{Vasilopoulos20}, and 1A\,0538-66 \citep{White78}, have shown occasional outbursts exceeding $L_{\rm X}\approx 10^{39}$~erg~s$^{-1}$, along with the first Galactic PULX, Swift~J0243.6+6124, discovered in 2017 \citep[][and references therein]{Wilson-Hodge18}.
Challenges in detecting PULXs and other transient bright accreting pulsars arise from their intermittent pulsations, low pulsed fractions, and transient nature. This suggests that the true population of PULXs, both in external galaxies and potentially in the Milky Way, is underestimated \citep[e.g.][]{Pintore25, Rodriguez20}.
Binary evolution models also make highly uncertain predictions for the number of ULXs with neutron stars per galaxy, with estimates depending on star formation history, metallicity, and magnetic field properties \citep{Shao15, Wiktorowicz21, Kuranov20, Kovlakas25, King23}. For example, some models suggest that only a handful of these binary systems exist in galaxies similar to the Milky Way \citep[e.g.][]{Kuranov20}.

To assess neutrino observability from transient Galactic PULXs, we adopted a hypothetical source with properties similar to those of Swift~J0243.6+6124. This source reached peak X-ray luminosities of $L_{\rm X}>10^{39}$~erg~s$^{-1}$ over 20 days \citep{Doroshenko20} and is located at $\sim5-7$~kpc \citep{Reig20}. No neutrino detections or upper limits for this source have been reported by IceCube (see Appendix \ref{swiftj0243 ul}). Given the transient nature of these systems, we followed the methodology of \citet{Abbasi22}, who compared neutrino fluence predictions for the 2015 V404~Cygni outburst (a non-PULX black hole system) to the $5\sigma$ discovery potential of IceCube.
In our analysis, for a fiducial PULX with $L_{\rm X}=5\times 10^{39}$~erg~s$^{-1}$
(see Table \ref{table simu} and Fig. \ref{spectra}), we re-scaled the IceCube discovery potential at 1~TeV from \citet{Abbasi22} to 100~GeV, accounting for the slope of the neutrino spectrum ($\gamma=-2.9$; see Table \ref{table simu}) and the detector energy-dependent effective area ratio $A_{\rm 1\,TeV}/A_{\rm 100\,GeV}\approx 400$ \citep{Stettner19}.
We adopted 100~GeV as the reference energy, rather than 1~TeV, to compare the expected neutrino fluence
with the $5\sigma$ discovery potential of IceCube, as this more effectively highlights the detection capabilities
for sources with soft neutrino spectra, as in our case.
Similarly to \citet{Anchordoqui03}, we assumed the phenomenological estimate for the beaming factor of $b=\Delta \Omega/4\pi\sim 0.1$ \citep{Cheng89}.
The resulting fluences, as a function of the time interval during which the source maintains an X-ray luminosity of $5\times 10^{39}$~erg~s$^{-1}$,
are shown in Fig. \ref{fluence} for three distances (2, 3, and 4~kpc), compared to the $5\sigma$ discovery potential.
We note that the extragalactic PULXs with X-ray luminosities exceeding those in our test case by a factor of $\sim 100$ have been observed \citep{Israel17b}. If such a luminous transient were to occur in the Milky Way, its neutrino emission would be detectable at distances well beyond the 2--4~kpc range explored here. Additionally, ongoing and future upgrades to DeepCore, the sub-array of IceCube sensitive to neutrinos in the $\sim10$~GeV to a few TeV energy range, will enhance its sensitivity. As shown in Fig. 3 of \citet{Abbasi22}, these advancements could lower the $5\sigma$ discovery potential by approximately a factor of 3, expanding detectability prospects (see also \citealt{Eller23}).

\section{Concluding remarks} \label{sect.conclusions}

We have shown that, within the framework of our model, nearby ULXs hosting neutron stars could be detectable sources
of neutrinos during periods of high X-ray activity ($L_{\rm x} > 10^{39}$~erg~s$^{-1}$), provided their X-ray outbursts last long enough.
Although no such sources have been observed nearby so far, their transient nature leaves open the possibility of a future discovery. The outlook for these studies is promising: upgrades of the currently active neutrino telescopes and new high-energy neutrino telescopes are either nearing completion or are planned for the next decade. Furthermore, combining data from these detectors is expected to significantly enhance sensitivity \citep{Schumacher25}, improving our ability to observe these sources.
We note a potential challenge for neutrino detection in luminous ULXs. High-resolution spectra of some ULXs, including pulsar-hosting systems, suggest outflows launched from the surface of the inner parts of the accretion disc. Such outflows may not always occur: for example, models indicate that outflows are expected to be negligible when the dipolar magnetic field is strong ($B>$few\,$10^{13}$~G;  \citealt{Mushtukov19}). If present, these outflows could affect the properties of the electrostatic gap, potentially suppressing it entirely. Crucially, the original gap models \citep{Cheng89, Cheng91} did not account for outflow effects.
We also note that our adopted weakly shielded gap assumption accounts in principle for the reduced potential drop across the gap caused by pair production from X-ray photons (compared to the strongly shielded gap case considered in most previous works). Nonetheless, the stronger disc radiation expected in ULXs could cause an additional reduction in this potential drop.
  Therefore, the formation and stability of gaps under these conditions remains uncertain and should be further investigated through observations and dedicated magnetohydrodynamic or particle-in-cell simulations.

\begin{acknowledgements}
  We thank the referee for the useful comments.
  LD acknowledges funding from the Deutsche Forschungsgemeinschaft (DFG, German Research Foundation) - Projektnummer 549824807.
  PR acknowledges support by the ``Programma di Ricerca Fondamentale INAF 2023''.
This research made use of NumPy \citep{Harris20}, SciPy \citep{2020SciPy-NMeth}, and pandas \citep{mckinney-proc-scipy-2010}.
\end{acknowledgements}

\bibliographystyle{aa} 
\bibliography{aa55242-25}

\begin{appendix}

\section{Flux upper limit determination for Swift~J0243.6+6124 in the IC86\_VII IceCube dataset} \label{swiftj0243 ul}

We used data analysis tools from the IceCube repository in {\tt Renkulab}\footnote{See \url{https://renkulab.io/projects/astronomy/mmoda/icecube} and the online data analysis (MMODA) service for IceCube, \url{https://mmoda.io}} and {\tt SkyLLH}\footnote{\url{https://icecube.github.io/skyllh/master/html/index.html}} to perform general maximum likelihood ratio hypothesis testing.
These tools do not allow arbitrary time intervals to be selected. Instead, they rely on predefined observation periods defined by the IceCube collaboration \citep{Abbasi21}.
Therefore, we analysed the IC86\_VII dataset, which roughly corresponds to the period of the Swift~J0243.6+6124 outburst.
From our analysis, we obtained a test statistic (TS) of 0.008, $n_{\rm s}=0.36$, and $\gamma=-5$.
Here, TS represents the likelihood ratio between the signal-plus-background and the background-only hypotheses, $n_{\rm s}$ is the best-fit
estimate of the number of signal events contributing to the data, and $\gamma$ characterizes the energy distribution of the signal.
We calculated the significance of the source, that is, the p-value, using the appropriate {\tt SkyLLH} tools.
This was done by generating the TS distribution of $10^5$ background-only data trials, which assumes zero injected signal events.
The resulting p-value is  0.47, meaning that there is not significant signal detection.
Assuming a spectral slope of $-3$ (that is, the slope obtained from our simulation for a source with $L_{\rm x}\approx 10^{38-39}$~erg~s$^{-1}$),
we derived a 90\% confidence level flux upper limit at 1~TeV of $5.7\times 10^{-11}$~TeV~cm$^{-2}$~s$^{-1}$ ($9.1\times 10^{-11}$~erg~cm$^{-2}$~s$^{-1}$).

\end{appendix}

\end{document}